\documentclass[usenatbib]{aa}

\usepackage{graphicx}
\usepackage{amssymb}
\usepackage{epsfig}
\usepackage{psfig}
\usepackage{natbib}
\def\pl{{\sc pl}}
\def\comptt{{\sc comptt}}

\def\bb{{\sc bb}}

\def\nh{{$N_{\rm H}$}}
\def\dbb{{\sc dbb}}
\def\bb{{\sc bb}}

\def\J1922{Swift~J1922.7-1716}
\def\be{\begin{equation}}
\def\ee{\end{equation}}

\begin{document}

\title{Swift, RXTE, and INTEGRAL observation of Swift~J1922.7-1716} 
\author{M. Falanga\inst{1,2}\fnmsep\thanks{\email{mfalanga@cea.fr}},
T. Belloni \inst{3},
S. Campana\inst{3}
}

\offprints{M. Falanga}
\titlerunning{Swift, RXTE, and INTEGRAL observation of Swift~J1922.7-1716}
\authorrunning{M. Falanga T. Belloni \& S. Campana}

\institute{CEA Saclay, DSM/DAPNIA/Service d'Astrophysique (CNRS FRE
  2591), F-91191, Gif sur Yvette, France
\and Unit\'e mixte de recherche Astroparticule et
Cosmologie, 11 place Berthelot, 75005 Paris, France
\and INAF - Osservatorio Astronomico di Brera, Via Bianchi 46, 23807
Merate, Italy 
}

\abstract{

We report the results of analyzing {\it Swift}, {\it RossiXTE},
and {\it INTEGRAL} data of \J1922, a likely transient X-ray source
discovered by Swift/BAT. Both the fast variability measured by the
{\it RXTE}/PCA and the combined {\it Swift}/XRT, {\it RXTE}/PCA, and
{\it INTEGRAL}/ISGRI (0.5--100 keV) energy spectrum suggest that the
system is a neutron-star low-mass X-ray binary or a black-hole candidate at low
accretion levels. The non-simultaneous spectra are consistent with the
same spectral shape and flux, suggesting little variability over the
period July to October 2005, but the analysis of archival {\it INTEGRAL}
data shows that the source was not detected in 2003--2004, suggesting
a transient or strongly variable behavior.


}
\maketitle

\section{Introduction}
\label{sec:intro}

\J1922\ was discovered  during the {\it Swift} 
Burst Alert Telescope (BAT) hard X-ray survey. The BAT survey covers
the 15--200 keV band and the time range December 2004 - March
2005 where \J1922\ was reported as $>5.5\sigma$ detection \citep[][]{tueller06a,tueller06b}. 
The BAT detection was also
confirmed in a follow-up observation with the {\it Swift} X-ray telescope
(XRT) in the 0.5--10 keV energy band. The source was found at the best-fit 
position $\alpha_{\rm J2000} = 19^{\rm h}22^{\rm m}37\fs0$ and
$\delta_{\rm J2000} = -17{\degr}17\arcmin02\farcs6$ with an estimated
uncertainty of $3\farcs2$ (90\% confidence). From the Ultra-Violet
Optical Telescope (UVOT) onboard {\it Swift}, the observed counterpart was not
consistent with the Palomar survey \citep{tueller06b}. 
Using the XRT data, the source spectrum was consistent with an absorbed
power-law model with a  photon index of $\Gamma=2.05\pm0.05$ and a
relatively low
equivalent hydrogen column density $N_H=(1\pm0.5)\times10^{21}$ cm$^{-2}$
\citep[][]{tueller06a,tueller06b}.    

\J1922\ was also observed as a target of opportunity (ToO) performed on October
21, 2005 with the {\it Rossi X-ray Timing Explorer} ({\em RXTE}), and the data
were made publicly available. The
source was also detected serendipitously during the {\it International
  Gamma-Ray Astrophysics Laboratory} ({\em INTEGRAL}) ToO observation of HETE
J1900.1-2455 performed from November 10--12, 2005.
In this letter we report the result of the spectral and timing
analysis of  \J1922\ using the {\it Swift}, {\em RXTE},
and {\em INTEGRAL} data. We analyze the broad band spectrum from 0.5--100
keV and perform a timing analysis to identify the nature of this new source.

\section{Observations and data}
\label{sec:observation}

\subsection{Swift}
\label{sec:SWIFR}
Swift carried out two observations of Swift J1922.7-1716 on July 8, 2005 and 
October 1, 2005. The XRT (Burrows et al. 2005) collected 6128 s data
and 8215 s data in photon counting mode, respectively. The high source-count
rate ($\sim 2$ cts s$^{-1}$) is such that the source is piled-up. We 
analyzed these data by extracting products from an annulus with an
inner radius of 5 pixels ($11.8''$) and outer radius 40 pixels
($94.3''$), where extracted 7468 counts and 10401 counts,
respectively. The background within the extraction region is
negligible ($<1\%$). 

For the spectral analysis, we generated appropriate ancillary response files with
the FTOOL task {\tt xrtmkarf}. Data were grouped to have 50 counts per
energy channel. The calibrated energy range is 0.5--10 keV using v.7 of
response files.

\subsection{RXTE}
\label{sec:RXTE}

We used publicly-available data from the proportional counter array
(PCA; 2--60 keV) \citep{jahoda96} and the High Energy X-ray Timing 
Experiment (HEXTE; 15--250 keV) \citep{rothschild98} onboard the {\it RXTE}
satellite. \J1922\ was observed on October 21, 2005
for two satellite orbits, from 08:54 UT for a total exposure time of 6.2 ks.  
For the spectral analysis, we extracted the PCA and HEXTE
energy spectrum using the standard software package FTOOLS version 6.0.2. 
However, the HEXTE detection, with a count rate of $\sim$ 1.82 cts/s/cluster,
was inadequate in this case for yielding a useful  high-energy
spectrum; therefore, it was excluded from the analysis. 
For the PCA, we limited the spectral and timing analysis using only the
PCU2 data, which was on during the whole observation.

\subsection{INTEGRAL}
\label{sec:INTEGRAL}

The present data was obtained during the
{\em INTEGRAL}  \citep{w03} ToO observation starting on October 27, 2005 and
ending on October 29, for a total exposure time of 210 ks.
The observation, aimed at  HETE J1900.1-2455, consists of 64 stable
pointings with a source position offset between $7^{\circ}$ and $11^{\circ}$.  
We used data from the coded mask imager IBIS/ISGRI \citep{u03,lebr03}
in the  20 to 200 keV energy range.  
The JEM-X monitor \citep{lund03} was not used since the
source was outside the field of view for all pointings. 
Data reduction was performed using the standard Offline Science 
Analysis (OSA) version 5.1.
The algorithms used for the spatial and spectral analyses are
described in \citet{gold03}. The ISGRI light curves are based on
events selected according to the detector 
illumination pattern  for \J1922. We used an illumination factor threshold
of 0.6 for the energy range 18--40 keV.  
In addition, we analyzed publicly-available ISGRI data from March
2003 to October 2004, for a total exposure time of 420 ks. In these data the
source was not detected at a statistically significant level in the
20--60 keV energy band.

\section{Results}
\label{sec:res}

\subsection{ISGRI imaging and light curve}
\label{sec:imaging}

In order to study the {\it INTEGRAL} light curve and spectrum of \J1922\ we
first deconvolved and analyzed the 64 single pointings separately and
then combined them into a total mosaic image in the 20--40 keV energy
band. In the mosaic, \J1922\ is clearly detected at a significance
level of 16$\sigma$. The source was detected with
the imaging procedure at the best-fit position $\alpha_{\rm J2000} =
19^{\rm h}22^{\rm m}36\fs2$ and $\delta_{\rm J2000} =
-17{\degr}16\arcmin18\farcs7$. This position, offset with respect
to the {\it Swift}/XRT positions \citep[][]{tueller06a,tueller06b}, is
$0\farcm44$. The 90\% confidence error on the ISGRI coordinates 
\citep[see ][]{gros03} is $1\farcm1$.
The background-subtracted 20--40 keV band light curve was extracted from
the images using all available pointings, each with a $\sim 3.3$ ks
exposure. The source mean-count rate was almost constant at $\sim 0.9$
cts s$^{-1}$ ($\sim 5.6\times10^{-11}$ erg cm$^{-2}$ s$^{-1}$). The count
rates are converted to flux using the spectrum model 4 described in
Sect. \ref{sec:spec}.  

\subsection{Timing analysis}
\label{sec:timing}

We searched for coherent pulsations over a wide range of periods
(2ms--1000s) in the total PCA energy range without finding any
significant signal. We derived the upper limits on the pulsed fraction
for a sinusoidal modulation (semi-amplitude of modulation divided by the
mean source count rate) at a 3$\sigma$ confidence level
\citep[see][for details of the algorithm used]{israel96}. For periods
longer than about 10s, we are not sensitive to any pulsations due to
the intrinsic aperiodic variability of the source. Upper limits in the
10\%--4\%, 3\%--4\%, and 4\%--5\% have been obtained for the 7s--1s,
1s--2.5ms, and 2.5ms--2ms period intervals, respectively. Also, when using the longer
observation but low-statistics ISGRI events data in the 18--40 keV
band, no coherent pulsation, orbital period or type-I X-ray bursts were found.

\begin{figure}[htb]
\centerline{\epsfig{file=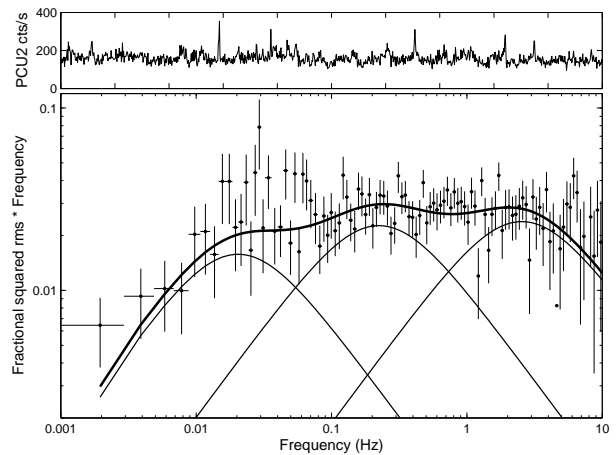,width=9.0cm}}
\caption
{Upper panel: the {\it RXTE}/PCA light curve of \J1922
  in the 3--15 keV energy band observed from 53664.37 to 53664.44 MJD.
  Lower panel: the  power density spectrum  in
  $\nu P_{\nu}$ form. The best fit Lorentzian models are also shown.} 
\label{fig:RXTE_timing}
\end{figure}

The 3--15 keV PCA light curve shows strong variability, around
$1.1\times10^{-10}$ erg cm$^{-2}$ s$^{-1}$, in the form of fast
flares with an excursion of about a factor of three, with a typical
duration of a few seconds (see upper panel of Fig. \ref{fig:RXTE_timing}). 
We studied the power density spectrum (PDS) of \J1922\ using the
high time-resolution PCA data. The PDS can be fitted with three
zero-centered Lorentzian components, for a total fractional rms variability,
integrated in the 0.001-10 Hz band, of $\sim$37\%.
The characteristic frequencies of the three Lorentzian
\citep[see][]{bpv02} are 0.042$\pm$0.009 Hz, 0.55$\pm$0.07   
Hz, and 5.5$\pm$1.0 Hz, about one decade apart from each other (see
Fig. 1). The high level of variability and the
Lorentzian decomposition are similar to those observed in black-hole
candidates (BHC) in their low/hard state and in a neutron-star (NS) 
low-mass X-ray binary system (LMXB) in their low-luminosity states
\citep[see e.g.,][]{b02}.

\subsection{Spectral analysis}
\label{sec:spec}

The spectral analysis was done using XSPEC version 11.3
\citep{arnaud96} for the 0.5--7 keV {\it Swift}/XRT data,
the 3--22 keV {\it RXTE}/PCA, and 20--100 keV
{\it INTEGRAL}/ISGRI data. For the broad-band spectral analysis, a
multiplicative factor for each instrument was included in the fit
to take into account the uncertainty in the cross-calibration of the
instruments, as well as variability across the non-simultaneous observations. 
The factor was fixed to 1 for the PCA data. All spectral
uncertainties in the results are given at a 90\% confidence level for
single parameters. 

First we fit the single data for each observation with a power-law (\pl) model 
including for the Swift/XRT data a photoelectrically-absorbed blackbody (\bb) 
model. We found that the spectral parameters are to the same order
within the error range (see Table \ref{table:spec1}). 
We then fit the 0.5--100 keV  broad-band spectrum using a simple
photoelectrically-absorbed \pl, model plus a \bb, model for thermal
soft emission below 3 keV and obtained a good  
$\chi^{2}{\rm /d.o.f.} = 243/207$. The best fit was found by
replacing the \pl\ with a cutoff \pl\ model obtaining an $\chi^{2}{\rm
  /d.o.f.} = 219/206$. However, in our case, the \pl\
  multiplied by a high-energy exponential cutoff model   
provides a slightly better description of the data compared to the   
\pl\ on the entire dataset at a 78\% confidence level (estimated by means of
an F-test). The best-fit values are found for a \bb\
temperature, $kT_{\rm soft}$, of $\sim 0.42$ keV, a \pl\ photon index of
$\Gamma\sim1.53$, and a cutoff energy of $\sim33$ keV. The interstellar
column density, $N_{\rm H}$, was found to be $\sim0.14\times10^{22}$
cm$^{-2}$. This value is within the same order as the
Galactic value, $N_{\rm H}=0.12\times10^{22}$ cm$^{-2}$, reported in
the radio maps of \citep{dickey90}.   

Assuming that the \pl\ is due to Comptonization
of soft photons by high-energy  electrons, we replaced the phenomenological cutoff
\pl\ model with a more physical thermal Comptonization model. For this,
we used the \comptt\ model  \citep{titarchuk94}, which is an  analytic model 
describing the Comptonization of soft photons described by a Wien spectrum
up-scattered in a hot plasma. The main
model parameters are the Thomson optical depth $\tau_{\rm T}$ across
the spherical or slab geometries, the electron temperature, $kT_{\rm  e}$, and 
the soft seed photon temperature, $kT_{\rm seed}$. The soft
thermal emission, $kT_{\rm soft}$ is fitted by a simple \bb\ or a
multi-temperature disc blackbody (\dbb) model \citep{mitsuda84}. This model
provides a good description of the data and the best-fit parameters are
reported in Table \ref{table:spec}.  

\begin{figure}[htb]
\centerline{\epsfig{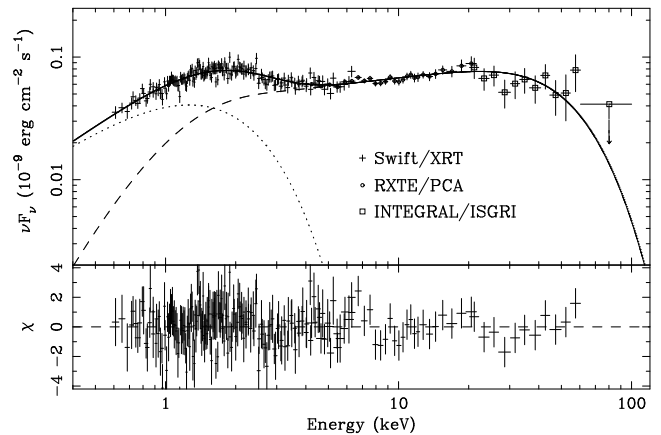}}
\caption
{The unfolded spectrum of \J1922\ fitted with an absorbed disc
  blackbody, \dbb, plus {\sc comptt} model. The data points correspond
  to the two XRT (0.5--7 keV), the PCA (4--22 keV), and the ISGRI (20--100
  keV) spectra, respectively. The \dbb\ model is shown by a dotted
  curve, the dashed curve gives the \comptt\ model, while the total
  spectrum is shown by a solid curve. The lower panel presents the
  residuals between the data and the model.}
\label{fig:spec}
\end{figure}

We found that the \dbb\ temperature, $kT_{\rm soft}$, and the soft seed photon
temperature, $kT_{\rm seed}$, are within the same order, suggesting that the
observed soft component is also the source of the seed
photons. Therefore, we repeated the fit with $kT_{\rm
  soft}=kT_{\rm seed}$ and obtained only a marginally worse $\chi^{2}{\rm
  /d.o.f.} = 226/206$. In Fig. \ref{fig:spec}, we show the unfolded
spectrum and the residuals of the data to the \dbb\ plus \comptt\ model. 
For the \comptt\ model, assuming a spherical geometry instead of a
slab/disc geometry leads to the same fit result. We tested for the
presence of iron emission line, but neither a 6.4
keV iron line nor Compton reflection were significantly detected.
The upper limit to the equivalent width for the iron line between 6.4 keV and 
6.9 keV is 176 eV (90\% confidence).
For all the fits, the normalizations of the XRT, PCA, and ISGRI data were
within $1.08\pm0.03$. Since \J1922\ was observed for {\it
 Swift}, {\it RXTE}, and {\it INTEGRAL} during different epochs, this 
 probably indicates
 that the source flux did not vary across the different observations.

\begin{table}[htb]
\caption{Spectra fit for the single  XRT, PCA, and ISGRI data.}
\begin{center}
\begin{tabular}{lllll}
\hline
Data set    & XRT 1             & XRT 2             & PCA      & ISGRI\\
Energy (keV) & 0.5--7 & 0.5--7 & 3--22 & 22--100\\
Parameters  & {\sc bb}+{\sc pl} & {\sc bb}+{\sc pl} & {\sc pl} & {\sc pl}\\
\hline
\noalign{\smallskip}
$N_{\rm H}$ $(10^{21} {\rm cm}^{-2})$   & 1.2$^{+0.5}_{-0.4}$ &
1.9$^{+0.5}_{-0.5}$ & -- & --\\ 
$kT_{\rm soft}$ (keV) & 0.43$^{+0.02}_{-0.02}$  & 0.40$^{+0.03}_{-0.03}$
 & -- & --\\
$R_{\rm BB}$ (km) & 13.4$\pm 0.4$ & 13.5$^{+0.1}_{-0.1}$ & -- & --\\
$\Gamma$& 1.63$^{+0.3}_{-0.4}$ & 1.75$^{+0.3}_{-0.4}$ & 1.83$^{+0.04}_{-0.04}$ & 2.2$^{+0.5}_{-0.6}$\\
$\chi^{2}/{\rm dof}$  &  89/88 & 83/82 &  21/25 & 10/9\\  
$F_{\rm X}^{(a)}$ & 1.7 & 1.8 & 1.1 & 0.8\\ 
\hline
\end{tabular}
\end{center}
{\noindent \small $^{(a)}$ unabsorbed flux in unit of $10^{-10}$erg cm$^{-1}$ s$^{-1}$.}
\label{table:spec1}
\end{table}

\begin{table*}[htb]
\caption{Best-fit parameters of the XRT/PCA/ISGRI data.}
\begin{center}
\begin{tabular}{lllllll}
\hline
          & Model 1 & Model 2 & Model 3 & Model 4 & Model 5 & Model 6\\
Parameters & {\sc bb}+{\sc pl}  & {\sc bb}+{\sc cutoff pl} & {\sc dbb}+{\sc cutoff pl} & {\sc bb}+\comptt & {\sc dbb}+\comptt & {\sc dbb}+\comptt\\
\hline
\noalign{\smallskip}
$N_{\rm H}$ $(10^{21} {\rm cm}^{-2})$  & 1.9$^{+0.4}_{-0.3}$ &
          1.4$^{+0.2}_{-0.4}$ & 2.0$^{+0.2}_{-0.2}$ & 1.5$^{+0.2}_{-0.2}$ &
            1.4$^{+0.3}_{-0.2}$ &  1.8$^{+0.3}_{-0.2}$\\ 
$kT_{\rm soft}$ (keV) & 0.41$^{+0.02}_{-0.02}$ &
          0.42$^{+0.01}_{-0.03}$  & 0.60$^{+0.03}_{-0.03}$ & 0.42$^{+0.02}_{-0.02}$  & 0.53$^{+0.05}_{-0.04}$ & 0.50$^{+0.03}_{-0.03}$\\
$R_{\rm in}\sqrt{\cos \, i}$ (km)$^{(a)}$ &-- &--& 5.7$^{+0.7}_{-0.5}$& -- &
            6.7$^{+1.8}_{-1.3}$ & 8.9$\pm 1.1$\\
$R_{\rm BB}$ (km) & 12.1$\pm 0.9$ & 12.5$\pm 1.0$ & -- &
            15.7$^{+1.5}_{-1.2}$ & -- & --\\
$\Gamma$& 1.8$^{+0.4}_{-0.4}$ & 1.43$^{+0.13}_{-0.15}$ &
          1.43$^{+0.13}_{-0.15}$ & -- & -- & --\\ 
$E_{\rm cutoff}$ (keV) &--& 32.5$^{+17}_{-7}$ & 28$^{+15}_{-8}$ & -- &
          -- & --\\$kT_{\rm seed}$ (keV) &-- & -- & -- &
          0.84$^{+0.12}_{-0.16}$ &  0.37$^{+0.08}_{-0.16}$ & $=kT_{\rm soft}$\\
$kT_{\rm e}$ (keV) &--& -- & --& 10.7$^{+3.2}_{-2.0}$ &
            10.7$^{+3.3}_{-2.1}$ & 10.6$^{+3.3}_{-2.1}$\\ 
$\tau_{\rm T}$ &--& -- & --& 3.04$^{+0.46}_{-0.54}$&
            3.04$^{+0.47}_{-0.52}$ & 3.06$^{+0.48}_{-0.51}$\\ 
$\chi^{2}/{\rm dof}$  &  243/207 &  219/206 & 232/206 &  211/205 & 219/205 & 226/206\\  
$F_{\rm 0.1-100keV}^{(b)}$  & 3.6$\times10^{-10}$ & 3.7$\times10^{-10}$ & 3.9$\times10^{-10}$&
            3.3$\times10^{-10}$ & 3.6$\times10^{-10}$ & 3.8$\times10^{-10}$\\ 
\hline
\end{tabular}
\end{center}
{\noindent \small $^{(a)}$ Assuming a distance of 10 kpc; $^{(b)}$
  unabsorbed flux in unit of erg cm$^{-1}$ s$^{-1}$.}
\label{table:spec}
\end{table*}

\section{Discussion and conclusions} 
 \label{sec:discussion}

The broad-band spectrum (0.5--100 keV) allowed us to perform an
improved spectral analysis for the new source \J1922\ using the {\it
Swift}/XRT, {\it RXTE}/PCA, and  {\it INTEGRAL}/ISGRI data.
The best fit to the data required a two-component
model, a cutoff \pl, or a thermal Comptonization model, together with
a soft component, see Table \ref{table:spec}. The hard spectral
  component contributes most of the observed flux (76\%), even though
  a soft \bb\ component is needed by the data.
Observationally, the
best-fit parameters are similar to those observed in BHC in their low/hard state  \citep[see e.g.][]{wilms06}
and in weakly magnetic neutron-star
LMXB in their low-luminosity states \citep[see e.g.][]{barret00}. 
In the low/hard or low-luminosity state hard X-ray
components extending up to energies of a few hundred keV have been clearly
detected in these systems. 
In NS systems, the hard spectrum is dominated by a \pl-like component, with a typical slope of $\Gamma \sim1.5-2.5$,
which is followed by an exponential cutoff at energies often $\gtrsim$
20 keV \citep[see e.g.,][]{barret00}, 
with some contribution from an additional soft thermal component,
$kT_{\rm soft}$.  
In BH systems in their low/hard state, the slope is around  $\Gamma \sim 1.5-1.6$,
with a high-energy cutoff of a few dozen keV \citep[see e.g.,][]{wilms06}, while
contribution from a soft thermal component is observable only when the
interstellar absorption is not too high \citep[see e.g.,][]{frontera01}.
The soft thermal emission, $kT_{\rm soft}$, could be associated to the
radiation from the accretion disc either for a NS of BHC source. 
The \pl\ is usually interpreted as Comptonization of seed photons
in hot, optical-depth plasma.  
Using the \comptt\ model, the hard spectrum is described by unsaturated
Comptonization of soft seed photons, $kT_{\rm seed}\sim 0.4$ keV, in the hot
$kT_{\rm e} \sim 11$ keV optically-thick $\tau \sim 3$ plasma. 

An important parameter is the distance to the source. Its galactic 
coordinates are $l_{II}$=20.7, $b_{II}$=-14.5, therefore in the
direction of the galactic bulge and substantially below the galactic
plane. Were the source within the galactic plane, its distance would be rather
small ($\sim1.5$ kpc for a thin disc); 
the derived absorption of $(1-2)\times 10^{21}$cm$^{-2}$ is compatible
with the total galactic absorption in that direction as estimated from
HI maps. Assuming a distance to the galactic center of 8 kpc and  
a bulge radius of 3 kpc, we conclude that the distance to \J1922\ is between
5 and 11 kpc. With these distance estimates, the unabsorbed 0.1--100 keV flux
of (3.3--3.9)$\times 10^{-10}$erg cm$^{-2}$ s$^{-1}$ (considering the
different models) translates to a luminosity of (1--5)$\times
10^{36}$ erg s$^{-1}$, compatible with both NS and BH sources at low
accretion-rate levels. We assume the source is probably located in the
galactic bulge. At the minimal distance of 5 kpc, the \bb\ fits
would imply an emission radius of only 3--4 km. This hints at a larger
source distance. The radii obtained with the \dbb\ fits are
also very small and depend crucially on the inclination of the source.  
Using a mean inclination angle of 60$^{\circ}$ and a source
distance of 10 kpc, an acceptable $R_{\rm in}\sim 13-18$ km can be
obtained. We note that the distance estimates assume that the
observed emission originates from the entire \bb\ surface facing
the observer. The presence of an obscuring structure, such as an
accretion disc or stream, will affect those estimates.

We conclude that we observed \J1922\ in its hard state, during which
it emits hard X-rays up to 100 keV.  The soft excess, $kT_{\rm
  soft}\sim 0.4$ keV detected at low energies is most likely
originating in the accretion disc. As for the hard component, it
most likely originates in the Comptonization of soft 
seed photons, $kT_{\rm seed}\sim 0.53$ keV in a hot plasma. Our results
are compatible with the soft emission being the source of the seed
photons, i.e. $kT_{\rm soft}=kT_{\rm seed}$.
From the best-fit spectral parameters and the timing properties, we cannot
tell whether the objects harbors a NS or a BH. 
Interestingly, we could obtain a satisfactory fit with the same model as all 
three non-simultaneous spectra ({\it Swift}/XRT, {\it
  RXTE}/PCA, and {\it INTEGRAL}/ISGRI), indicating that its spectrum most
likely remained constant if the source varied between the observations. Once again, this is a typical
observational fact both in NS and BH systems at low luminosity.
The measured \nh\ value is consistent with the expected Galactic value
in the source direction \J1922. This consistency  suggests the absence
of  intrinsic absorption and, together with the position in the sky, indicates
that it is most likely located in the Galactic bulge. The lack of iron-line
emission is consistent with the absence of a reflected component in
the broad band spectrum of the source. The fact that \J1922\ was not
 detected in the ISGRI data from March 
2003 to October 2004, for a total 20--60 keV exposure time of 420 ks,
indicates that the system is either very variable on long time scales
or transient. Across the period spanned by our observations (2005 July
to October), both flux level and spectrum were compatible with being constant.
 
\acknowledgements
We are grateful to G. L. Israel for support for the search for
coherent pulsations. MF acknowledge the French Space Agency (CNES)
for financial support. TB and SG acknowledge support from ASI grants 
I/R/046/04 and I/023/05/0.


\begin{thebibliography}{99} 
 
\bibitem[\protect\citeauthoryear{Arnaud}{1996}]{arnaud96} 
Arnaud, K. A. 1996, in Astronomical Data 
Analysis Software and Systems V. 
ed.\ G. H. Jacoby, \& J. Barnes,   ASP Conf. Series 101 (San 
Francisco: ASP), 17 

\bibitem[\protect\citeauthoryear{Barret et al.}{2000}]{barret00}
    Barret, D., Olive, J. F., Boirin, L., et al. 2000, ApJ, 533, 329

\bibitem[\protect\citeauthoryear{Belloni, Psaltis \& van der
    Klis}{2002}]{bpv02} Belloni, T., Psaltis, D., \& van der Klis,
    M. 2002, ApJ, 572, 392

\bibitem[\protect\citeauthoryear{Belloni et al.}{2002}]{b02} Belloni,
  T., Colombo, A. P., Homan, J., et al. 2002, A\&A, 390, 199 

\bibitem[\protect\citeauthoryear{Dickey \& Lockman}{1990}]{dickey90}
  Dickey, J. M., \& Lockman, F. J. 1990, Annu. Rev. A\&A, 28, 215
  
\bibitem[\protect\citeauthoryear{Frontera et al.}{2001}]{frontera01}
Frontera, F., Zdziarski, A. A., Amati, L., et al., 2001, ApJ, 561, 1006

\bibitem[\protect\citeauthoryear{Goldwurm et al.}{2003}]{gold03} 
Goldwurm, A., Daid, P.,  Foschini, L., et al. 2003, A\&A, 411, L223 
 
\bibitem[\protect\citeauthoryear{Gros et al.}{2003}]{gros03} 
Gros, A., Goldwurm, A., Cadolle-Bel,  M., et al. 2003, A\&A, 411, L179 

\bibitem[\protect\citeauthoryear{Israel \& Stella}{1996}]{israel96} 
Israel, G. L., Stella, L. 1996, ApJ, 468, 369

\bibitem[\protect\citeauthoryear{Jahoda et al.}{1996}]{jahoda96} 
  Jahoda, K., Swank, J. H., Giles, A. B., et al. 1996, Proc. SPIE, 2808, 59 
 
\bibitem[\protect\citeauthoryear{Lebrun et al.}{2003}]{lebr03} 
Lebrun,  F., Leray, J.-P., Lavocate, Ph., et al. 2003, A\&A, 411, L141 

\bibitem[\protect\citeauthoryear{Lund et al.}{2003}]{lund03} 
Lund, N., Budtz-Joergensen, C.,  Westgaard, N. J., et al.  2003, A\&A,
411, L231  

\bibitem[\protect\citeauthoryear{Mitsuda et al.}{1984}]{mitsuda84} 
Mitsuda, K., Inoue, H.,  Koyama K., et al. 1984, PASJ, 36, 741 

\bibitem[\protect\citeauthoryear{Rothschild et al.}{1998}]{rothschild98} 
Rothschild, R. E., Blanco, P. R., Gruber, D. E., et al. 1998, ApJ, 496, 538 
 
\bibitem[\protect\citeauthoryear{Titarchuk} {1994}]{titarchuk94} 
Titarchuk L., 1994, ApJ, 434, 570  

\bibitem[\protect\citeauthoryear{Tueller et al.} {2006a}]{tueller06a} 
Tueller, J., Barthelmy, S., Burrows, D., et al. 2006, Astr. Tel., 668

\bibitem[\protect\citeauthoryear{Tueller et al.} {2006b}]{tueller06b} 
Tueller, J., Barthelmy, S., Burrows, D., et al. 2006, Astr. Tel., 669

\bibitem[\protect\citeauthoryear{Ubertini et al.}{2003}]{u03} 
Ubertini, P., Lebrun, F., Di  Cocco, G., et al. 2003, A\&A, 411, L131 

\bibitem[\protect\citeauthoryear{Wilms et al.}{2006}]{wilms06} 
Wilms, J., Nowak, M.A., Pottschmidt, K., Pooley, G.G., Fritz, S., 2006, A\&A, 447, 245
 
\bibitem[\protect\citeauthoryear{Winkler et al.}{2003}]{w03} 
Winkler, C., Courvoisier, T. J.-L., Di  Cocco, G., et al. 2003, A\&A, 411, L1 
 
\end{thebibliography}
\end{document}